%% file: main.tex
\documentclass[11pt,letterpaper]{article}
\usepackage[letterpaper,top=1.0in,bottom=1.0in,left=1.0in,right=1.0in]{geometry}
\usepackage{url}
\usepackage{times}
\usepackage{color}
\usepackage{subfigure}
\usepackage{paralist}
\usepackage{wrapfig}
\usepackage{authblk}

\usepackage{pgfplots}

\newcommand{\D}{\ensuremath{\mathcal{D}}}

\newcommand{\xor}{\ensuremath{\oplus}}

\newcommand{\W}{\ensuremath{\mathcal{W}}}
\newcommand{\K}{\ensuremath{\mathcal{K}}}
\newcommand{\WM}{\ensuremath{\mathcal{WM}}}

\newcommand{\sysname}{{\sc Lime}}

\renewcommand\paragraph[1]{\smallskip \noindent {\bf #1.}}

\begin{document}
\title{Lime: Data Lineage in the Malicious Environment
}
\pagenumbering{arabic}
\author{Michael Backes}
\author{Niklas Grimm}
\author{Aniket Kate}
\affil{Saarland University, Germany\\
  \texttt{\{backes@cs, s9nigrim@stud, aniket@mmci\}.uni-saarland.de}}

\date{}

\maketitle
\begin{abstract}
\vspace{-0ex}
Intentional or unintentional leakage of confidential data is undoubtedly one of the most severe security
threats that organizations face in the digital era.
The threat now extends to our personal lives:
a plethora of personal information is available to social networks and smartphone providers
and is indirectly transferred to untrustworthy third party and fourth party applications.

In this work, we present a generic data lineage framework \sysname\ for data flow across multiple entities
that take two characteristic, principal roles (i.e., owner and consumer).
We define the exact security guarantees required by such a data lineage mechanism toward identification of a guilty entity, and
identify the simplifying non-repudiation and honesty assumptions. % necessary .
We then develop and analyze a novel accountable data transfer protocol between two entities
within a malicious environment by building upon oblivious transfer, robust watermarking, and signature primitives.
Finally, we perform an experimental evaluation to demonstrate the practicality of our protocol.

\end{abstract}
\input{intro}

\input{sysmod}
\input{definitions}

\input{untrusted_sender}

\input{implementation}

\input{scenarios}

\input{discussion}

\input{relwork}

\input{conclusion}

\paragraph{Acknowledgements}
We thank Deepak Garg for encouraging discussions and Amit Roy, Emre Goynugur  and Isaac Alpizar Chacon 
for their work towards an initial draft. We are also grateful to 
our anonymous reviewers for their comments.

{
\footnotesize
\raggedright
\bibliographystyle{abbrv}
\bibliography{main}
}

\end{document}

%% file: intro.tex
\section{Introduction}\label{sec:Intro}
In the digital era, information leakage through unintentional exposures,
or intentional sabotage by disgruntled employees and malicious external entities,
present one of the most serious threats to organizations. % and individuals alike.
According to an interesting chronology of data breaches maintained by 
the Privacy Rights Clearinghouse (PRC), in the United States alone,
$867,525,654$ records have been breached from $4,289$ data breaches made public since $2005$~\cite{DataBreach}.
It is not hard to believe that this is just the tip of the iceberg, as most cases of information leakage 
go unreported due to fear of loss of customer confidence or regulatory penalties:
it costs companies on average \$$214$ per compromised record~\cite{DataBreachCost}.
Large amounts of digital data can be copied at almost no cost and can be spread through the internet in very short time. Additionally, the risk of getting caught for data leakage is very low, as there are currently almost no accountability mechanisms. For these reasons, the problem of data leakage has reached a new dimension nowadays.

Not only companies are affected by data leakage, it is also a concern to individuals. 
The rise of social networks and smartphones has made the situation worse.
In these environments, individuals disclose their personal information to various service providers,
commonly known as {\em third party applications}, in return for some possibly free services.
In the absence of proper regulations and accountability mechanisms, 
many of these applications share individuals' identifying information 
with dozens of advertising and Internet tracking companies. 

Even with access control mechanisms, where access to sensitive data is limited, a malicious authorized user can publish sensitive data as soon as he receives it. Primitives like encryption offer protection only as long as the information of interest is encrypted, but once the recipient decrypts a message, nothing can prevent him from publishing the decrypted content. Thus it seems impossible to prevent data leakage proactively.

Privacy, consumer rights, and advocacy organizations such as
PRC~\cite{prc} and EPIC~\cite{epic} try to address the problem of information leakages 
through policies and awareness.
However, as seen in the following scenarios
the effectiveness of policies is questionable
as long as it is not possible to provably associate the guilty parties to the leakages. 

\paragraph{Scenario 1: Social Networking}
It was reported that third party applications of the widely used online social network Facebook leak sensitive private information about the users or even their friends to advertising companies~\cite{FacebookAppLeak}.
In this case, it was possible to determine that several applications were leaking data by analyzing their behaviour and so these applications could be disabled by Facebook. However, it is not possible to make a particular application responsible for leakages that already happened, as many different applications had access to the private data.

\paragraph{Scenario 2: Outsourcing} 
Up to $108,000$ Florida state employees were informed that their personal information has been compromised due to improper outsourcing ~\cite{offshoreOutsourcing}. The outsourcing company that was handed sensitive data hired a further subcontractor that hired another subcontractor in India itself. Although the offshore subcontractor is suspected, it is not possible to provably associate one of the three companies to the leakage, as each of them had access to the data and could have possibly leaked it.

\smallskip

We find that the above and other data leakage scenarios 
can be associated to an absence of accountability mechanisms during data transfers:
leakers either do not focus on protection, or they intentionally expose confidential data without any concern, 
as they are convinced that the leaked data cannot be linked to them.
In other words, when entities know that they can be held accountable for leakage of some information,
they will demonstrate a better commitment towards its required protection. 

In some cases, identification of the leaker is made possible by forensic techniques, but these are usually expensive and do not always generate the desired results.
Therefore, we point out  the need for a general accountability mechanism in data transfers.
This accountability can be directly associated with \emph{provably} detecting a 
transmission history of data across multiple entities starting from its origin.
This is known as data provenance, data lineage or source tracing.
The data provenance methodology,
in the form of robust watermarking techniques~\cite{MSU06} 
or adding fake data~\cite{papadimitriou2011data}, 
has already been suggested in the literature 
and employed by some industries.
However, most efforts have been ad-hoc in nature and there is no formal model available. Additionally, most of these approaches only allow identification of the leaker in a non-provable manner, which is not sufficient in many cases.

\paragraph{Our Contributions}
In this paper, we formalize this problem of provably associating the guilty party to the leakages, 
and work on the data lineage methodologies to solve the problem of
information leakage in various leakage scenarios.

As our first contribution, we define \sysname, a generic data lineage framework
for data flow across multiple entities in the malicious environment.
We observe that entities in data flows assume one of two roles: 
owner or consumer. 
We introduce an additional role in the form of auditor, 
whose task is to determine a guilty party for any data leak,
and define the exact properties for communication between these roles.
In the process, we identify an optional non-repudiation assumption made between  
two owners, and an optional trust (honesty) assumption made by the auditor about the owners.

The key advantage of our model is that it enforces {\em accountability by design}; i.e.,
it drives the system designer to consider possible data leakages and the corresponding accountability 
constraints at the design stage. This helps to overcome the existing situation 
where most lineage mechanisms are applied only after a leakage has happened.

As our second contribution, we present an accountable data transfer protocol 
to verifiably transfer data between two entities.
To deal with an untrusted sender and an untrusted receiver scenario associated with data transfer between two consumers,
our protocols employ an interesting combination of the robust watermarking, oblivious transfer, and signature primitives.
We also implement our protocol and demonstrate the practicality for real-life data transfer scenarios such as online social networks and outsourcing.

\paragraph{Paper Outline}
We organize the remainder of this paper as follows: In Section~\ref{sec:sysmod} we introduce our model \sysname\ and give a threat model and design goals. Section~\ref{definitions} describes primitives used in the paper and Section~\ref{transfer_protocols} presents and analyzes protocols for accountable data transfer. Section~\ref{implementaion} shows results we obtained from a practical implementation 
 and Section~\ref{sec:scenarios} gives examples of how our model can be applied to real world settings.
We discuss additional features of our approach in Section~\ref{discussion} and related work in Section~\ref{sec:relwork}. Finally we present our conclusions in Section~\ref{sec:conclusion}.

%% file: sysmod.tex
\section{The Lime Framework}\label{sec:sysmod}
As we want to address a general case of data leakage in data transfer settings, we propose the simplifying model \sysname\ (\textbf{L}ineage \textbf{i}n the \textbf{m}alicious \textbf{e}nvironment). With \sysname\ we assign a clearly defined role to each involved party and define the inter-relationships between these roles. This allows us to define the exact properties that our transfer protocol has to fulfill in order to allow a provable identification of the guilty party in case of data leakage.

\subsection{Model}
As \sysname\ is a general model and should be applicable to all cases, we abstract the data type and call every data item \emph{document}. There are three different roles that can be assigned to the involved parties in \sysname: data \emph{owner}, data \emph{consumer} and \emph{auditor}.
The data owner is responsible for the management of documents and the consumer receives documents and can carry out some task using them. The auditor is not involved in the transfer of documents, he is only invoked when a leakage occurs and then performs all steps that are necessary to identify the leaker. All of the mentioned roles can have multiple instantiations when our model is applied to a concrete setting. We refer to a concrete instantiation of our model as \emph{scenario}. 

In typical scenarios the owner transfers documents to consumers. However, it is also possible that consumers pass on documents to other consumers or that owners exchange documents with each other.
In the outsourcing scenario~\cite{offshoreOutsourcing} the employees and their employer are owners, while the outsourcing companies are untrusted consumers.

In the following we show relations between the different entities and introduce optional trust assumptions. 
We only use these trust assumptions because we find that they are realistic in a real world scenario and because it allows us to have a more efficient data transfer in our framework. At the end of this section we explain how our framework can be applied without any trust assumptions.

When documents are transferred from one owner to another one, we can assume that the transfer is governed by a \emph{non-repudiation assumption}. 
This means that the sending owner trusts the receiving owner to take responsibility if he should leak the document. As we consider consumers as untrusted participants in our model, a transfer involving a consumer cannot be based on a non-repudiation assumption.
Therefore, whenever a document is transferred to a consumer, the sender embeds information that uniquely identifies the recipient. We call this \emph{fingerprinting}. If the consumer leaks this document, it is possible to identify him with the help of the embedded information. 

As presented, \sysname\ relies on a technique for embedding identifiers into documents, as this provides an instrument to identify consumers that are responsible for data leakage. We require that the embedding does not not affect the utility of the document. Furthermore, it should not be possible for a malicious consumer to remove the embedded information without rendering the document useless. A technique that can offer these properties is \emph{robust watermarking}. We give a definition of watermarking and a detailed description of the desired properties in  Section~\ref{DefWM}.

\begin{figure}[!t]
\centering
	 \includegraphics[width=.6\textwidth]{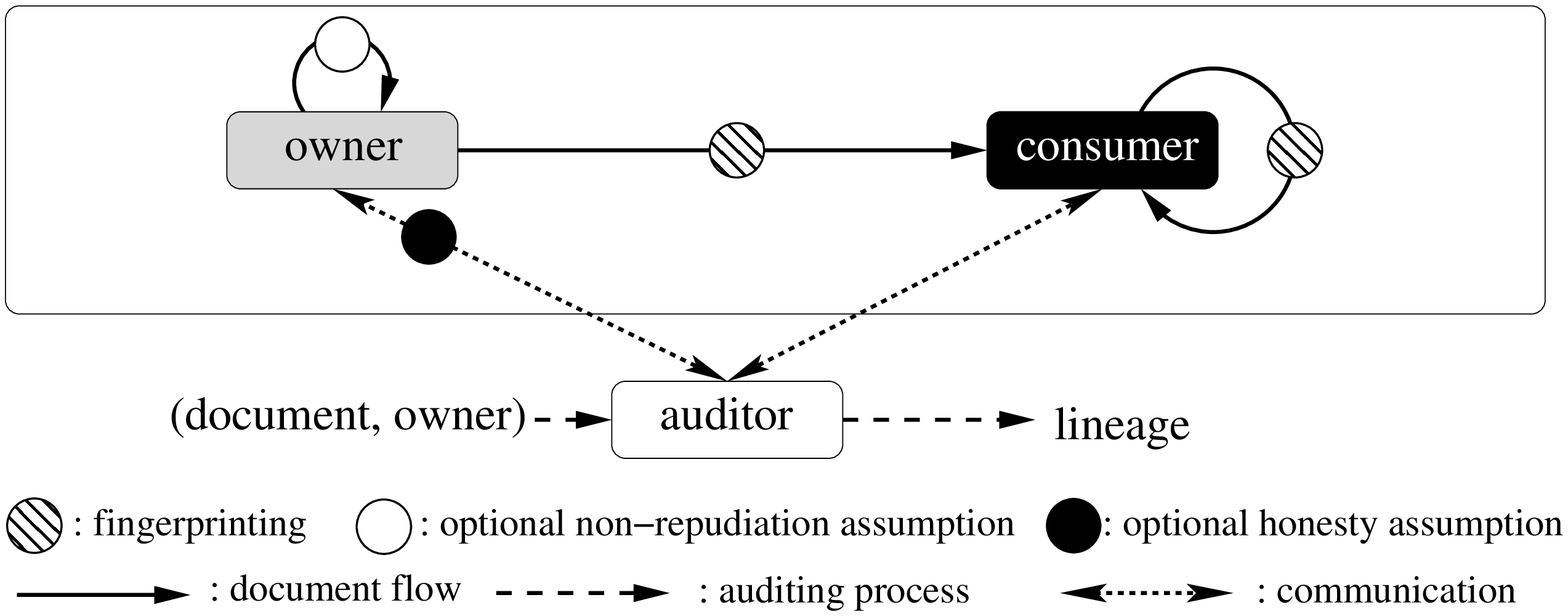}
	 \caption{\label{figure_model}The \sysname\ framework}
\end{figure}

A key position in \sysname\ is taken by the auditor. He is not involved in the transfer, but he takes action once a leakage occurs. He is invoked by an owner and provided with the leaked data. If the leaked data was transferred using our model, there is identifying information embedded for each consumer who received it. Using this information the auditor can create an ordered chain of consumers who received the document. We call this chain the \emph{lineage} of the leaked document. The last consumer in the lineage is the leaker.  In the process of creating the lineage  each consumer can reveal new embedded information to the auditor to point to the next consumer -- and to prove his own innocence. 
In order to create a complete lineage it is necessary that the auditor receives information from the owner, as only the owner can reveal the information embedded during the first transfer. We assume that the auditor is always invoked by the owner or that he is at least provided with information about the owners identity, so that the auditor can start his investigation with the owner and  a complete lineage can be created. 

We can assume that the auditor trusts the owner to be honest. Honesty in this case means, that the owner does not leak a document and blame another party. We can make this assumption 
as the owner is concerned with the document's privacy in the first place. However, the auditor does not trust the consumers. 
In a real world setting the auditor can be any authority, for example a governmental institution, police, a legal person or even some software. 
In the outsourcing scenario~\cite{offshoreOutsourcing}, the employer can invoke the auditor who recreates the lineage and thereby uncovers the identity of the leaker. The employer can use this information to take legal actions against the outsourcing company.
We show the flow of documents, the optional non-repudiation and honesty assumptions  the cases in which fingerprinting is used in Figure~\ref{figure_model}.
In Section~\ref{sec:scenarios}, we present how \sysname\ can be applied to the outsourcing scenario presented in the introduction.

\paragraph{Remark}
We choose these honesty and non-repudiation assumptions because they realistically model many real-world scenarios. Due to these assumptions we can reduce the overhead introduced to transfers by \sysname: In a transfer between two owners no fingerprinting has to be used and the owner can run a simplified transfer protocol because he is trusted by the auditor. However, these trust assumptions are not necessary for the correctness of \sysname. If an owner is untrusted we can easily treat him as a consumer, so we do not have to make any trust assumptions about him.

\subsection{Threat Model and Design Goals}
\label{threat_model}

Although we try to address the problem of data leakage, \sysname\ cannot guarantee that data leakage does not occur in the first place; once a consumer has received a document, nothing can prevent him from publishing it. We offer a method to provably identify the guilty party once a leakage has been detected. By introducing this \emph{reactive} accountability, we expect that leakage is going to occur less often, since the identification of the guilty party will in most cases lead to negative consequences. As our only goal is to identify guilty parties, the attacks we are concerned about are those that disable the auditor from provably identifying the guilty party. 

Therefore, we  consider attackers in our model as consumers that take every possible step to publish a document without being held accountable for their actions. As the owner does not trust the consumer, he uses fingerprinting every time he passes a document to a consumer. However, we assume that the consumer tries to remove this identifying information in order to be able to publish the document safely. 
As already mentioned previously, consumers might transfer a document to another consumer, so we also have to consider the case of an \emph{untrusted sender}. This is problematic because a sending consumer who embeds an identifier and sends the marked version to the receiving consumer could keep a copy of this version, publish it and so frame the receiving consumer. Another possibility to frame other consumers is to use fingerprinting on a document without even performing a transfer and publish the resulting document. 

A different problem that arises with the possibility of false accusation is denial. If false accusation is possible, then every guilty receiving consumer can claim that he is innocent and was framed by the sending consumer.

The crucial phase in our model is the transfer of a document involving untrusted entities, so we clearly define which properties we require our protocol to fulfill. We call the two parties \emph{sender} and \emph{recipient}.
We expect a transfer protocol %in the untrusted environment 
to fulfill the following properties and only tolerate failures with negligible probabilities.
\begin{compactenum}
	\item {\bf correctness:} When both parties follow the protocol steps correctly and only publish their version of the document, the guilty party can be found.
  \item {\bf no framing:} The sender cannot frame recipients for the sender's leakages.
	\item {\bf no denial:} If the recipient leaks a document, he can be provably associated with it.
\end{compactenum}

We also require our model to be collusion resistant, i.e. it should be able to tolerate a small number of colluding attackers~\cite{PW97}.
We also assume that the communication links between parties are reliable.

\paragraph{Non-Goals}
We do not aim at proactively stopping data leakage, we only provide means to provably identify the guilty party in case a leak should occur, so that further steps can be taken.
We also do not aim for integrity, as at any point an entity can decide to exchange the document to be sent with another one. However, in our settings, the sender wants the receiver to have the correct document, as he expects the recipient to perform a task using the document so that he eventually obtains a meaningful result.

Our approach does not account for derived data (derived data can for example be generated by applying aggregate functions or other statistical operations), as much of the original information can be lost during the creation process of derived data.
Nevertheless, we show in Section~\ref{composability} how \sysname\ can operate on composed data. We think of composed data as a form of data created from multiple single documents, so that the original documents can be completely extracted (e.g. concatenation of documents).

\label{fairness} We do not consider {\em fairness} issues in our accountable transfer protocol; more precisely, 
we do not consider scenarios in which %a consumer requests the transfer of a certain document and 
a sender starts to run the transfer protocol but aborts before a recipient received the document,
or when a recipient, despite of receiving the document, 
falsely claims that he did not receive it.
In real-world scenarios, we find fairness not to be an issue 
as senders and recipients expect some utility from the transfer, 
and are worried about their reputation and corporate liabilities.

%% file: definitions.tex
\section{Primitives}
\label{definitions}
A function $f$ is \emph{negligible} if for all $c > 0$ there is a $n_c$ so that for all $n \ge n_c$ $f(n) \le \frac{1}{n^c}$.
In our scheme we make use of \emph{digital signatures}. More precisely, we use a CMA-secure signature~\cite{CMA}, i.e., no polynomial-time adversary is able to forge a signature with non-negligible probability. For a message $m$ that has been signed with party $A$'s signing key, we write $[m]_{sk_A}.$
We use a symmetric encryption scheme that offers security under chosen plaintext attacks, writing $c = enc(m,ek)$ for encryption of a message $m$ and $m = dec(c,ek)$ for decryption of a ciphertext $c$ with symmetric key $ek$.

\subsection{Robust Watermarking}
\label{DefWM}
We use the definition of watermarking by Adelsbach et al.\ \cite{AKS07}.
To argue about watermarking, we need a so-called similarity function $sim(D,D')$ that returns $\top$ if the two documents $D$ and $D'$ are considered similar in the used context and $\bot$ otherwise. The similarity function is a different one for each data type used and we assume it is given.

Let \D~be the set of all possible documents, $\WM \subseteq \{0,1\}^+$ the set of all possible watermarks, \K~the set of keys and $\kappa$ the security parameter of the watermarking scheme.
A \emph{symmetric, detecting watermarking scheme} is defined by three polynomial-time algorithms:
\begin{asparaitem}
  \item The probabilistic \emph{Key Generation Algorithm} $GenKey^{WM}(1^\kappa)$ outputs a key $k \in \K$ for a given security parameter $\kappa$. 
  \item The probabilistic \emph{Embedding Algorithm} generates a watermarked document $D' = \W(D, w, k)$ on input of the original document $D \in \D$, the watermark $w \in \WM$ and the key $k \in \K$.
  \item The \emph{Detection Algorithm} $Detect(D', w, D, k)$ outputs $\top$ or $\bot$ on input of a (potentially watermarked) document $D' \in \D$, a watermark $w \in \WM$, the original document $D \in \D$  and a key $k \in \K$. $\top$ means that the watermark is detectable; $\bot$ means, that it is not.
\end{asparaitem}

We require the following properties:
\begin{asparaitem}
	\item {\bf imperceptibility: }$\forall D \in \D, \forall w \in \WM, \forall k \in \K. D' \leftarrow \W(D, w, k) \Rightarrow sim(D,$ $  D') = \top$, i.e., the original document and the watermarked document are similar.
	\item {\bf effectiveness: }  $\forall D \in \D, \forall w \in \WM, \forall k \in \K.D' \leftarrow \W(D, w, k) \Rightarrow Detect(D',$ $ w, D, k) = \top$, i.e., if a watermark is embedded using a key $k$, the same watermark should be detectable using the same key $k$.
\item {\bf robustness: }For a watermarked document $D' = \W(D, w, k) , D\in\D, w\in\WM, k\in\K$ there is no polynomial-time adversary that can compute a $D''\in \D$ given $D'$ and $w$ so that $sim(D',D'') = \top$ and $Detect(D'', w, D,  k) = \bot$ with non-negligible probability. This means that no adversary can efficiently remove or change a watermark without rendering the document unusable.
\end{asparaitem}

Additionally, we require our watermarking scheme to support \emph{multiple} \emph{re-water-} \emph{marking}, i.e., it should allow for multiple (bounded by the dataflow path length) watermarks to be embedded successively without influencing their individual detectability. This property can also be considered as a special kind of robustness, as it prevents adversaries from making a watermark undetectable simply by adding more watermarks using the same algorithm. More information and some experimental results about this property can be found in \cite{MSU06}.

We also expect the watermarking scheme to be \emph{collusion resistant}~\cite{kilian1998resistance}, i.e., even if an attacker can obtain differently watermarked versions of a document, he should not be able to create a version of the document were none of these watermarks is detectable. 
Further, for some watermarking schemes the input of the original document is not required for detection of watermarks. We call those watermarking schemes \emph{blind}.
As already stated in \cite{AKS07} this definition of watermarking is very strong and its properties are not yet provided by available schemes. Although we chose this strong definition to prove the correctness of our scheme, there are existing schemes whose properties are arguably sufficient in practice such as the Cox watermarking scheme~\cite{cox1997secure}.

\paragraph{Cox Watermarking Scheme~\cite{cox1997secure}}
This scheme uses spread-spectrum embedding of Gaussian noise to watermark images. To provide robustness, the watermark is embedded in the most significant part of the picture, so that removing the watermark should not be possible without destroying the underlying picture. The $\alpha$-factor of the algorithm is a parameter that determines how strong the Gaussian noise is influencing the original image, thus it can be used to trade robustness against imperceptibility. Although our strong robustness requirement is not formally fulfilled, it is shown that the scheme is robust against many common attacks such as scaling, JPEG compression, printing, xeroxing and scanning, multiple re-watermarking and others~\cite{cox1997secure}.

\subsection{1-out-of-2 Oblivious Transfer}
1-out-of-2 Oblivious Transfer $(OT_1^2)$ involves two parties, the {\it sender} and the {\it chooser}. The sender offers two items $M_0$ and $M_1$ and the chooser chooses a bit $\sigma$. The chooser obtains $M_{\sigma}$ but no information about $M_{1-\sigma}$ and the sender learns nothing regarding $\sigma$. In this context, when speaking of learning nothing, we actually mean nothing can be learned with non-negligible probability.
 There are different concrete instantiations of this primitive.

\begin{figure}[!t]
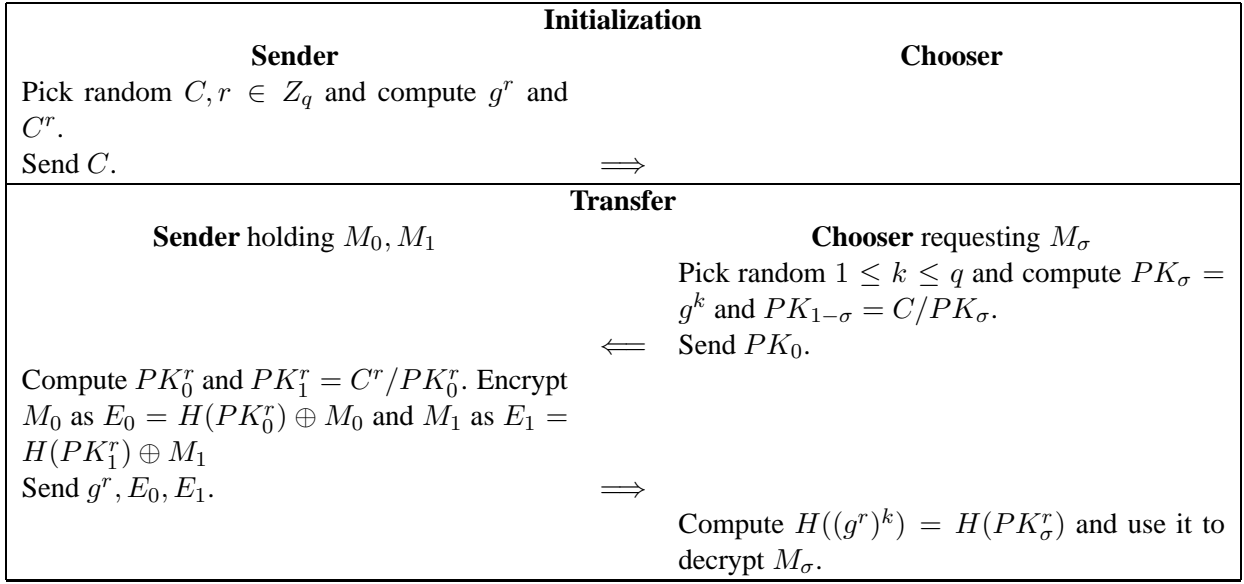

\begin{tabular}{| p{.44\textwidth}   c   p{.44\textwidth} |}
	\hline
	\multicolumn{3}{|c|}{\bf Initialization} \\	
	\multicolumn{1}{|c}{\bf Sender} & &  \multicolumn{1}{c|}{\bf Chooser} \\
	Pick random $C, r \in Z_q$ and compute $g^r$ and $C^r$.	& & \\
	Send $C$.	&	$\Longrightarrow$	& \\
	\hline
	
	\multicolumn{3}{|c|}{\bf Transfer} \\	
	\multicolumn{1}{|c}{{\bf Sender} holding $M_0, M_1$} & &  \multicolumn{1}{c|}{{\bf Chooser} requesting $M_\sigma$} \\
	&	&	Pick random $1 \leq k \leq q$ and compute $PK_{\sigma} = g^k$ and $PK_{1-\sigma} = C / PK_{\sigma}$.\\
	& $\Longleftarrow$ & Send $PK_0$. \\
	Compute $PK_0^r$ and $PK_1^r = C^r/PK_0^r$. Encrypt $M_0$ as $E_0 = H(PK_0^r) \xor M_0$ and $M_1$ as $E_1 = H(PK_1^r) \xor M_1$ & & \\
	Send $g^r, E_0, E_1$.	& $\Longrightarrow$ & \\
	&	& Compute $H((g^r)^k) = H(PK_\sigma^r)$ and use it to decrypt $M_\sigma$. \\
	\hline
	
\end{tabular}
\caption{Oblivious Transfer protocol by Naor and Pinkas \cite{NP01}}
	\label{oblivious_transfer}
\end{figure}

As an example implementation of $OT^2_1$ we show a protocol by Naor and Pinkas ~\cite{NP01}:
The protocol operates over a group $Z_q$ of prime order $q$ with a generator $g$ for which the computational Diffie-Hellman assumption holds. The protocol is proven secure in the random oracle model using an ideal hash function $H$. The construction can be found in Figure~\ref{oblivious_transfer}.

In the initialization phase, the sender picks a random group element $C$ and sends it to the chooser. It is important that the chooser does not know $DLog(C)$. Additionally, the sender chooses another random group element $r$ and computes $C^r$. These values are not used in the initialization phase, but for efficiency reasons the sender can compute these values in this phase, as it can be performed offline.

In the transfer phase, the chooser picks a random $1\leq k \leq q$ and computes $PK_{\sigma}$ = $g^k$ and $PK_{1-\sigma} = C/PK_{\sigma}$, so that he knows $DLog(PK_{\sigma})$. Then he sends $PK_0$ to the sender. 
The sender computes $PK_0^r$ and without further exponentiation $PK_1^r = C^r / PK_0^r$. Then he encrypts the two messages $M_0$ and $M_1$ as $E_0 = H(PK_0^r) \xor M_0$ and $E_1 = H(PK_1^r) \xor M_1$ and sends $g^r, E_0$ and $E_1$ to the chooser.
The chooser can now compute $H((g^r)^k) = H(PK_{\sigma}^r)$ and so decrypt $M_{\sigma} = E_{\sigma} \xor H(PK_{\sigma}^r)$.

The proof for the correctness and security of the protocol in the random oracle model and under the Diffie-Hellman assumption can be found in \cite{NP01}. Additionally, they provide an approach to improve efficiency for multiple executions of $OT_1^2$ using a bandwidth/computation tradeoff. This is achieved by performing one single instance of $OT_1^{2^n}$ offering all $2^n$ possible combinations instead of $n$ instances of $OT_1^2$ offering every part on its own. Depending on available computational power and bandwidth, this can improve efficiency.

When we use $OT_1^2$ in our protocols to send messages, the sender actually encrypts the messages, sends both encryptions to the chooser and performs $OT_1^2$ just on the decryption keys. This allows us to use the $OT_1^2$ protocol with a fixed message size while actually sending messages of arbitrary size. Note that this could only be a security risk if the chooser was able to break the encryption scheme.

%% file: untrusted_sender.tex
\section{Accountable Data Transfer}
\label{transfer_protocols}

In this section we specify how one party transfers a document to another one, what information is embedded and which steps the auditor performs to find the guilty party in case of data leakage.
We assume a public key infrastructure to be present, i.e. both parties know each others signature verification key.

\subsection{Trusted Sender}
\begin{figure}[!t]
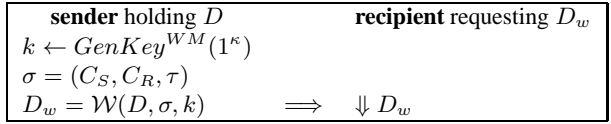

\centering
\footnotesize
\begin{tabular}{|l c l|}
	\hline
	\multicolumn{1}{|c}{\textbf{sender} holding $D$} 	&	& \multicolumn{1}{c|}{\textbf{recipient} requesting $D_w$} \\
  $k \leftarrow GenKey^{WM}(1^\kappa)$					&	& \\	
	$\sigma = (C_S, C_R, \tau)$ & & \\
	$D_w = \W(D,\sigma,k)$	&	$\Longrightarrow$ &$ \Downarrow D_w$ \\
	\hline
\end{tabular}
\caption{protocol for trusted senders}
\label{trusted_senders}
\end{figure}

In the case of a trusted sender it is sufficient for the sender to embed identifying information, so that the guilty party can be found. As the sender is trusted, there is no need for further security mechanisms.
In Figure~\ref{trusted_senders}, we present a transfer protocol that fulfills the properties of correctness and no denial as defined in Section~\ref{threat_model}. As the sender is trusted to be honest, we do not need the no framing property.

The sender, who is in possession of some document $D$, creates a watermarking key $k$, embeds a triple $\sigma = (C_S, C_R, \tau)$ consisting of the two parties' identifiers and a timestamp $\tau$ into $D$ to create $D_w = \W(D, \sigma, k)$. He then sends $D_w$ to the recipient, who will be held accountable for this version of the document. As the sender also knows $D_w$, this very simple protocol is only applicable if the sender is completely trusted; otherwise the sender could publish $D_w$ and blame the recipient.

\subsection{Untrusted Sender}
In the case of an untrusted sender we have to take additional actions to prevent the sender from cheating, i.e. we have to fulfill the no framing property. To achieve this property, the sender divides the original document into $n$ parts and for each part he creates two differently watermarked versions. He then transfers one of each of these two versions to the recipient via $OT_1^2$. The recipient is held accountable only for the document with the parts that he received, but the sender does not know which versions that are. The probability for the sender to cheat is therefore $\frac{1}{2^n}$. We show the protocol in Figure~\ref{complete_protocol} and  provide an analysis of the protocol properties in Section~\ref{analysis}.

First, the sender generates two watermarking keys $k_1$ and $k_2$. It is in his own interest that these keys are \emph{fresh} and \emph{distinct}. The identifying information that the sender embeds into the document $D$ is a signed statement $\sigma = [C_S,C_R,\tau]_{sk_{C_R}} $ containing the sender's and recipient's identifiers and a timestamp $\tau$, so that every valid watermark is authorized by the recipient. 
The sender computes the watermarked document $D'=\W(D,\sigma,k_1)$, splits the document $D'$ into $n$ parts and creates two different versions $D_{i,j}=\W(D_i,j,k_2)$ of each part by adding an additional watermark $j\in \{0,1\}$. 
For each version of each part $D_{i,j}$, $i\in\{1, \dots, n\}, j\in\{0,1\}$ he creates a signed message $m_{i,j} = [\tau,i,j]_{sk_{C_S}}$ containing the timestamp of the current transfer, the part's index and the content of the version's second watermark. Then he generates an AES key $ek_{i,j}$, encrypts $c_{i,j} = enc(\langle D_{i,j}, m_{i,j} \rangle, ek_{i,j})$ and sends $c_{i,j}$ to the recipient. The recipient chooses a bit $b_i \in \{0,1\}$ for each $i \in \{1 \dots n\}$ and receives $ek_{i,b_i}$ via oblivious transfer (note that the $n$ executions of $OT_2^1$ can be parallelized). He then decrypts all $c_{i,b_i}$ using $ek_{i,b_i}$ and reconstructs the document by joining the parts $D_{1,b_1} \dots D_{n,b_n}$. The signed statements $m_{1,b_1} \dots m_{n,b_n}$ serve as proof of his choice. As for each part he chooses a bit $b_i$, the final version is identified by a bitstring $\overline{b} \in \{0,1\}^n$. As he will only be held accountable for the version watermarked with $\overline{b}$, we  have a failure probability (i.e., the sender correctly guessing $\overline{b}$) of $\frac{1}{2^n}$.

\subsection{Data Lineage Generation}
The auditor is the entity that is used to find the guilty party in case of a leakage. He is invoked by the owner of the document and is provided with the leaked document. In order to find the guilty party, the auditor proceeds in the following way:

\begin{enumerate}
	\item The auditor initially takes the owner as the current suspect.
  \item \label{loop} The auditor appends the current suspect to the lineage.
  \item The auditor sends the leaked document to the current suspect and asks him to provide the detection keys $k_1$ and $k_2$ for the watermarks in this document as well as the watermark $\sigma$. If a non-blind watermarking scheme is used, the auditor additionally requests the unmarked version of the document.
  \item If, with key $k_1$, $\sigma$ cannot be detected, the auditor continues with \ref{end}.
	\item If the current suspect is trusted, the auditor checks that $\sigma$ is of the form $(C_S,C_R,$ $\tau)$ where $C_S$ is the identifier of the current suspect, takes $C_R$ as current suspect and continues with \ref{loop}.
	\item The auditor verifies that $\sigma$ is of the form $[C_S, C_R, \tau]_{sk_{C_R}}$ where $C_S$ is the identifier of the current suspect. He also verifies the validity of the signature.
  \item The auditor splits the document into $n$ parts and for each part he tries to detect $0$ and $1$ with key $k_2$. If none of these or both of these are detectable, he continues with \ref{end}. Otherwise he sets $b'_i$ as the detected bit for the $i$th part. He sets $\overline{b'} = b'_1 \dots b'_n$.
  \item The auditor asks $C_R$ to prove his choice of $\overline{b} = b_1 \cdots b_n$ for the given timestamp $\tau$ by presenting the $m_{i,b_i} = [\tau, i, b_i]_{sk_{C_S}}$. If $C_R$ is not able to give a correct proof (i.e., $m_{i,b_i}$ is of the wrong form or the signature is invalid) or if $\overline{b} = \overline{b'}$, then the auditor takes $C_R$ as current suspect and continues with \ref{loop}.
	\item \label{end} The auditor outputs the lineage. The last entry is responsible for the leakage.
\end{enumerate}

\paragraph{Remark}
It would also be possible to include the signed statement $\sigma$ in every single part of the document, but as the maximum size of a watermark is limited by the document's size, this might be problematic for scenarios where the parts are small. Therefore, we embed $\sigma$ in the complete original document and only embed single bits to the (possibly small) parts of the document.
We use a timestamp $\tau$ to uniquely identify a specific transfer between two parties, and thus assume that no two transfers between the same two parties take place at the same time. However, it would be possible to use a counter that is incremented on each transfer to allow multiple transfers at the exact same time.

\subsection{Analysis of the Protocol}
\label{analysis}

We now show that the protocol presented in Figure~\ref{complete_protocol} fulfills the required properties of correctness, no framing and no denial as presented in Section~\ref{threat_model} : 

\begin{figure*}[!tb]
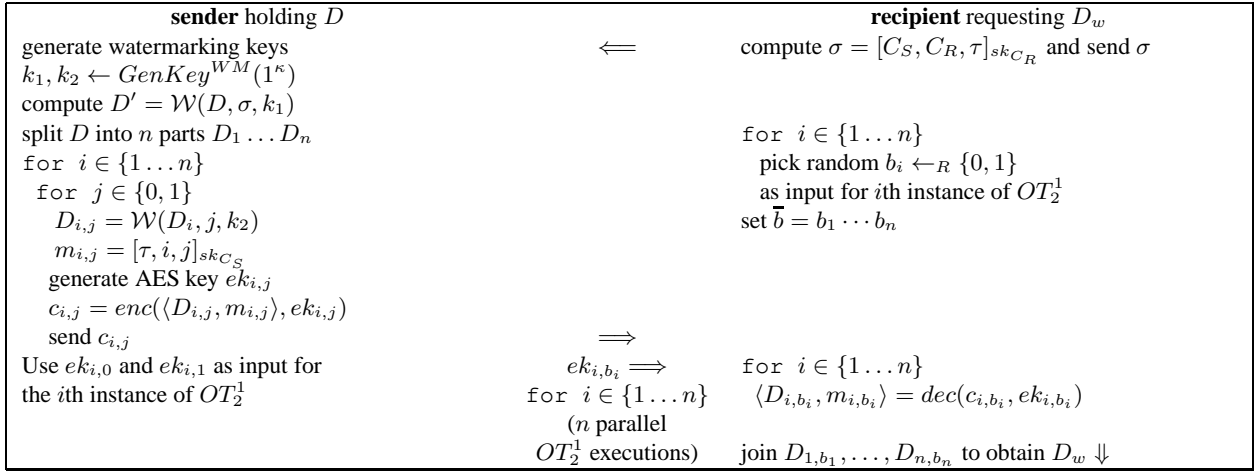

\footnotesize
\centering
\begin{tabular}{| p{.38\textwidth}  c  p{.40\textwidth} |}
	\hline
	\multicolumn{1}{|c}{\textbf{sender} holding $D$} 	&	& \multicolumn{1}{c|}{\textbf{recipient} requesting $D_w$} \\
  generate watermarking keys \newline 
  $k_1, k_2 \leftarrow GenKey^{WM}(1^\kappa)$ \newline
%  k_2 \leftarrow GenKey^{WM}(1^\kappa)$ \newline
	compute $D' = \W(D,\sigma,k_1)$	 &$\Longleftarrow$	& compute  $\sigma = [C_S,C_R,\tau]_{sk_{C_R}} $ and send $\sigma$ \\
	split $D$ into $n$ parts $D_1 \dots D_n$ \newline	
	$\texttt{for } i \in \{1 \dots n\}$ 	\newline
	\hspace*{5pt}$\texttt{for } j \in \{0,1\}$ \newline
	\hspace*{10pt} $D_{i,j} = \W(D_i,j,k_2)$ \newline
	\hspace*{10pt} $m_{i,j} = [\tau,i,j]_{sk_{C_S}}$ \newline
	\hspace*{10pt}generate AES key $ek_{i,j}$ \newline
	\hspace*{10pt}$c_{i,j} = %\newline  \hspace*{15pt} 
	enc(\langle D_{i,j}, m_{i,j} \rangle, ek_{i,j})$
	& &$\texttt{for } i \in \{1 \dots n\}$	\newline
			\hspace*{5pt} pick random $b_i \leftarrow_R \{0,1\}$ \newline
			\hspace*{5pt} as input for $i$th instance of $OT_2^1$ \newline
			set $\overline{b} = b_1 \cdots b_n$ \\		
	\hspace*{10pt}send $c_{i,j}$ & $\Longrightarrow$ & \\
      Use $ek_{i,0}$ and $ek_{i,1}$ as input for &	$ek_{i,b_i}$ $\Longrightarrow$ & $\texttt{for } i \in \{1 \dots n\} $ \\
      the $i$th instance of $OT_2^1$&	$\texttt{for } i \in \{1 \dots n\} $	 & \hspace*{5pt}$\langle D_{i,b_i}, m_{i, b_i} \rangle = dec(c_{i,b_i}, ek_{i,b_i})$\\
  & ($n$ parallel & \\%$\hspace*{10pt}  \\
  &	$OT_2^1$ executions) &	join $D_{1,b_1}, \dots , D_{n,b_n}$ to obtain $D_w \Downarrow$ \\
%	&	&	$\Downarrow D_w$ \\
	\hline
\end{tabular} 
	\caption{\label{complete_protocol}protocol for untrusted senders}

\end{figure*}

\begin{enumerate}
	\item {\bf correctness:} Assume that both parties follow the protocol steps correctly. We show that for all possible scenarios the guilty party is determined correctly:
		\begin{enumerate}	
			\item \emph{the sender publishes $D$ or $D'$}: The auditor does not detect $\sigma$ (in the case of $D$) or the $b_i$ (in the case of $D'$) and correctly blames the sender.
			\item \emph{the recipient publishes $D_w$}: The auditor successfully detects $\sigma$ and $\overline{b'}$ and verifies that $\sigma$ is of the correct form. The recipient is able to provide the proof of his choice of $\overline{b}$; the auditor verifies $\overline{b'} = \overline{b}$ and suspects the recipient. As there are no further watermarks embedded, the auditor correctly blames the recipient.
		\end{enumerate}
  \item {\bf no framing:} In a first step we show that the sender cannot obtain the version of the document for which the recipient can prove his choice (i.e. the version watermarked with the bitstring $\overline{b}$):
	   The sender knows all the $D_{i,j}$, that are used for the computation of $D_w$, but he does not know the bitstring $\overline{b}$ that the recipient chose due to the properties of oblivious transfer. So all he can do is guess a bitstring $\overline{b^*} = b^*_1 \cdots b^*_n$ and create $D_w^* = join (D_{1,b^*_1}, \dots, $ $ D_{n,b^*_n})$. But $\overline{b}$ and $\overline{b^*}$ (and therefore $D_w$ and $D_w^*$) are the same only with negligible probability of $\frac{1}{2^n}$, so the probability for the sender to learn $D_w$ is negligible if he follows the protocol correctly. \\
    The sender might try to learn about $\overline{b}$ by offering the same version twice during the oblivious transfer. Usually the recipient would have no possibility of realizing this, as he cannot detect the watermark, but as the sender additionally has to send the signed statements $m_{i,j} = [\tau,i,j]_{sk_{C_S}}$ the recipient knows that he received what he asked for. This is the case because $j$ has by construction the same value as $b_i$ chosen by the recipient. The sender can still send a wrong version $D_{i,1-j}$ and so know the bitstring the document is watermarked with, but as the recipient only proves his \emph{choice} of $b$, the sender still cannot frame him; he would only lose the ability to prove the recipients' guilt in case the recipient publishes the document.

   In a second step we show that a malicious sender cannot create a document that the recipient will be held accountable for without running the transfer protocol: \\
  As the correctness of the signed statement $\sigma$ is verified in the auditing process and as the sender can only forge the recipient's signature with negligible probability, the only possibility to mount this attack is to reuse a valid signed statement from a past transaction. This implies that the included timestamp $\tau$ is the same, too. As the auditor asks the recipient to prove his choice of $\overline{b}$ for this $\tau$, the recipient is able to provide a correct proof, as a valid transfer with timestamp $\tau$ actually happened. Analogous to the previous case, the sender can only chose $\overline{b^*} \in \{0,1\}^n$ randomly and therefore he can only succeed with negligible probability.

  From these two steps it follows that the sender is not able to frame a recipient.

  \item{\bf no denial:} We first show that the recipient cannot publish a version of the document whose embedded watermarks are different from those embedded in the version $D_w$ of the document that he righteously obtained. He also cannot obtain the watermark-free version $D$:
	As the recipient only receives the watermarked version, he could only learn $D$ by removing the watermark, which he can only do with negligible probability due to the robustness property of the watermarking scheme. The recipient could also create a watermarked version with a different bitstring embedded, if he is able to get $D_{i,1-b_i}$ for some $i$, but this is only possible if he breaks the $OT_2^1$ scheme or the encryption scheme, which is only possible with negligible probability.

  We now show that a recipient cannot cheat during the auditing process, when he proves which version of the document he asked for during the transfer protocol: 
In order to prove another choice of $\overline{b}$ he would again have to break the $OT_2^1$ scheme or the encryption scheme in order to learn the $m_{i,1-b_i}$ for some $i$ or he would need to forge the sender's signature to create $m_{i,1-b_i}$ for some $i$. As all of this is only possible with negligible probability, the overall probability for the recipient to succeed is negligible. 

From these two steps it follows that the recipient is not able to publish a document without being caught.
\end{enumerate}

This proves our protocol's security up to fairness issues. However, as discussed in Section~\ref{fairness}, 
we do not find fairness issues to be problematic in our application scenarios.

%% file: implementation.tex
\begin{figure}[!t]
  \subfigure[]{
	\includegraphics[width=.45\textwidth]{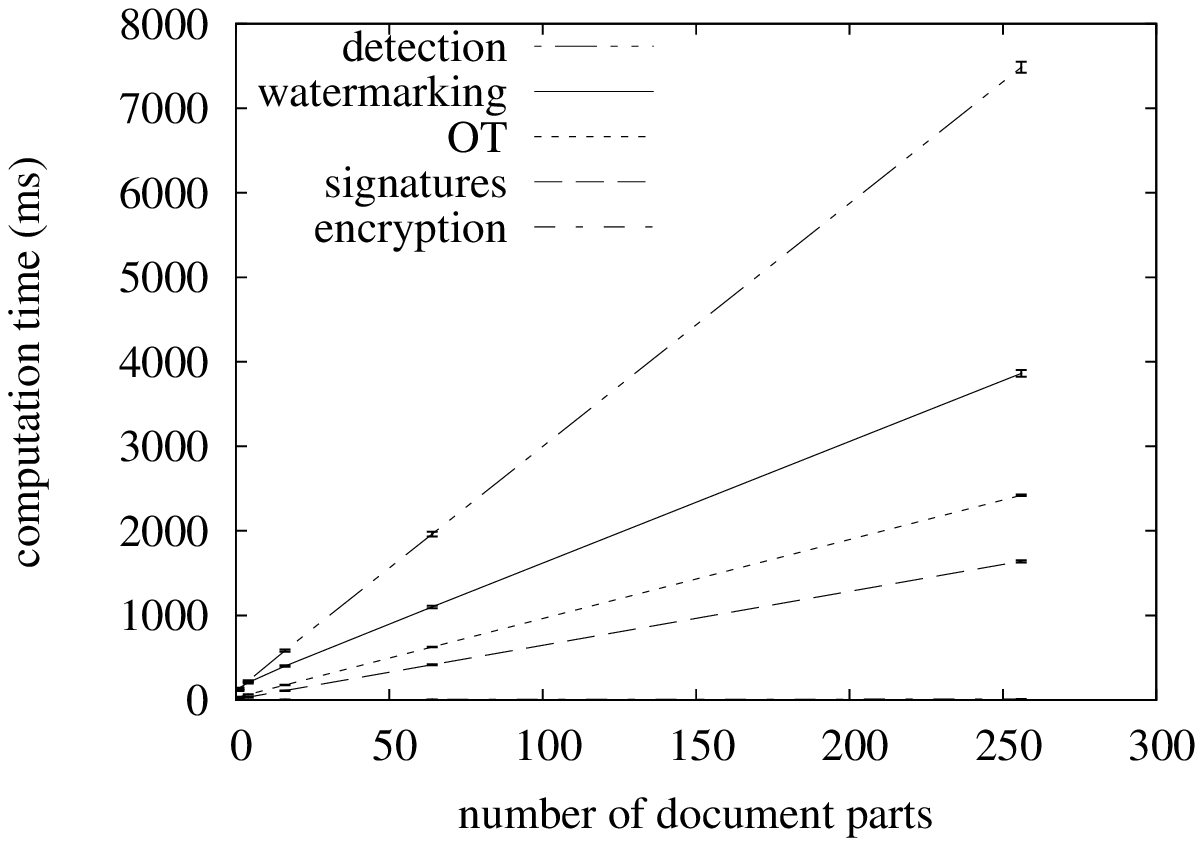}
  \label{experiment_1}
}
\subfigure[]{
	\includegraphics[width=.45\textwidth]{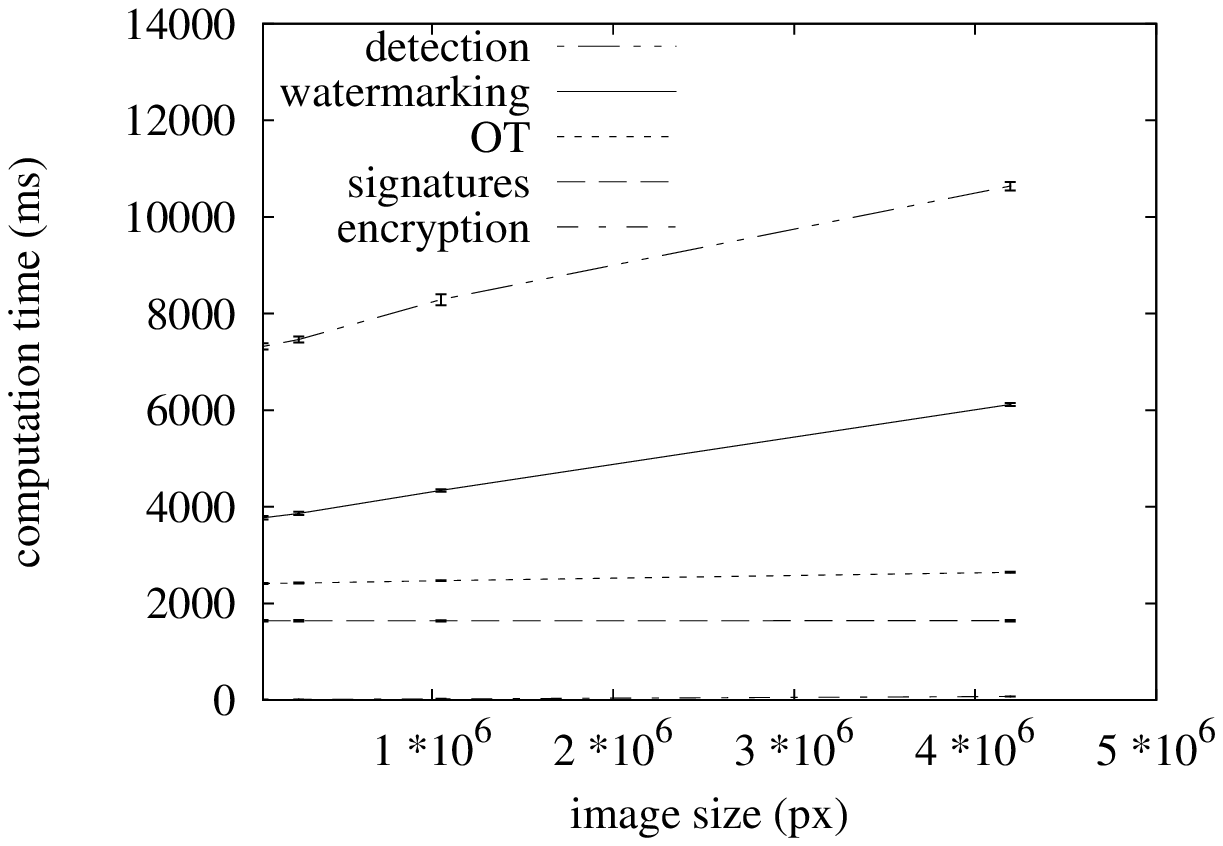}
  \label{experiment_2}
}
\caption{(a) shows computation times for different numbers of document parts; (b) shows computation times for different image sizes.}
\end{figure}

\section{Implementation and Microbenchmarking}
\label{implementaion}

We implemented the protocol in Figure~\ref{complete_protocol} as a proof-of-concept and to analyze its performance. For the oblivious transfer subprotocol we implemented the protocol by Naor and Pinkas~\cite{NP01} using the PBC library~\cite{PBClib}, which itself makes use of the GMP library~\cite{GMPlib}. For signatures we implemented the BLS scheme~\cite{boneh2001short}, also using the PBC library. For symmetric encryption we used an implementation of AES from the Crypto++~\cite{CryptoPPlib} library. For watermarking we used an implementation of the Cox algorithm for robust image watermarking~\cite{cox1997secure} from Peter Meerwald's watermarking toolbox \cite{watermarkingToolbox}. We set the $\alpha$-factor, which determines the strength of the watermark, to a value of 0.1.

We executed the experiment with different parameters to analyze the performance. The sender and recipient part of the protocol are both executed in the same program, i.e., we do not analyze network sending, but only computational performance. The executing machine is a Lenovo ThinkPad model T430 with 8 GB RAM and 4 $\times$ 2.6GHz cores, but all executions were performed sequentially. We measured execution times for different phases of the protocol: watermarking, signature creation, encryption, oblivious transfer and detection. We executed each experiment 250 times and determined the average computation time and the standard deviation.

\begin{figure}[!t]
  \center
  \subfigure[]{\includegraphics[width=.48\textwidth]{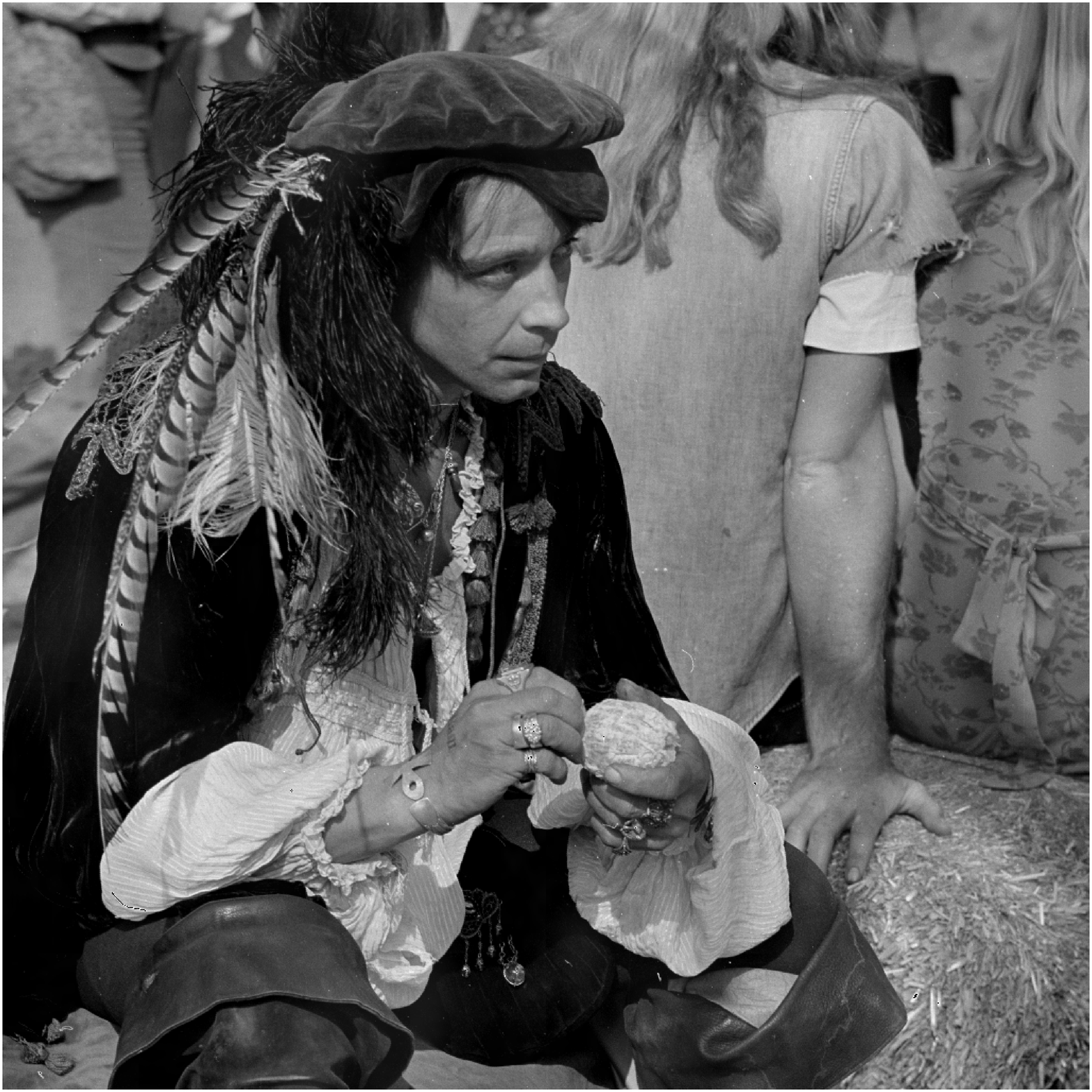}
  \label{lena_non_scrambled}
}
\subfigure[]{
	\includegraphics[width=.48\textwidth]{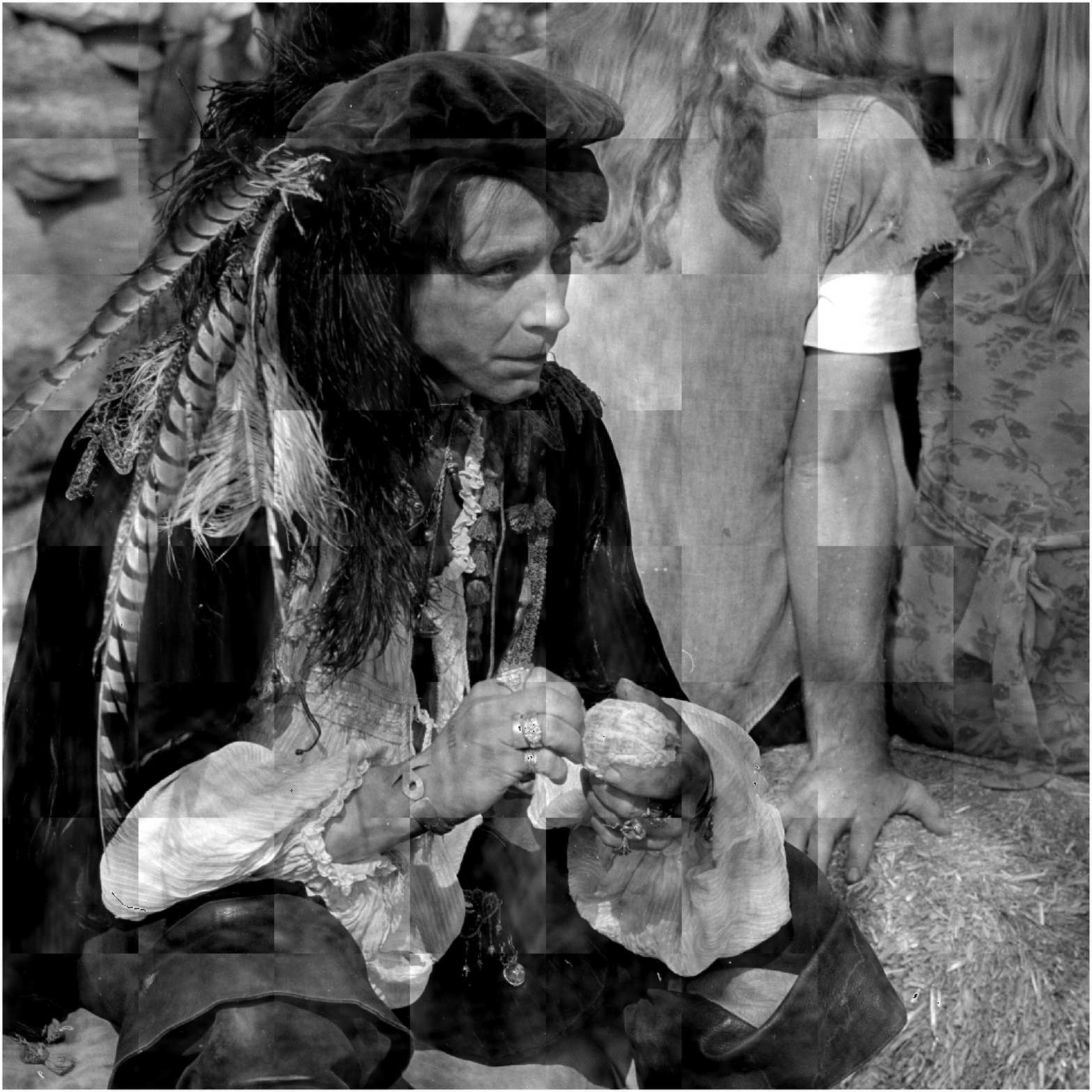}
  \label{lena_scrambled}}

  \caption{(a) shows an image transferred with our algorithm. In (b)  we set the $\alpha$-factor used by the Cox algorithm to 0.5 in order to obtain very strong watermarks for the small parts. As a result differences between adjacent document parts are visible.}
\end{figure}

In the first experiment we used an image of size $512 \times 512$ pixels and changed the number of parts the image was split into. We show the results in Figure~\ref{experiment_1}.
We can see that the execution time of watermarking, signatures, oblivious transfer and detection is linear in the number of document parts. The execution time of encryption is also increasing slowly, but it is still insignificant compared to the other phases. 

In the second experiment we used a fixed number of document parts of 256 and changed the size of the image used. We used 4 different sizes: $256 \times 256$ px (65KB), $512 \times 512$ px (257KB), $1024 \times 1024$ px (1.1 MB) and $2048 \times 2048$ px (4.1 MB). The results can be seen in Figure~\ref{experiment_2}.
We can observe that the execution time for watermarking and detection is linear in the image size, but in contrast to the previous result, we already start with a long execution time for small sizes and the growth is rather slow. The execution time for oblivious transfer stays constant, which might be surprising at first sight, but this can be explained easily, as we only transfer AES keys in the oblivious transfer phase and these are of constant size for all image sizes. Of course the encrypted files also have to be sent over the network (so there would be a higher communication overhead for bigger images), but in our implementation we only considered the computational costs. The execution time for the creation of the signatures is also constant as the number and form of the signed statements is the same for all images. Again the execution time of encryption is increasing slowly, but it is of no significance compared to the other phases.

The tests showed that from the resulting files all watermarks (i.e. the identifying watermark $\sigma$ and all the single bits $b_i$ forming the bitstring $\overline{b}$) could be correctly detected, showing the correctness of our protocol. 
In Figure~\ref{lena_non_scrambled} in the appendix we give an example of an image that was transferred using our algorithm. 

We find that latencies of a few seconds are acceptable in the scenarios that we considered. Additionally, as we show in our experiments in Figure~\ref{experiment_1}, one can easily perform a tradeoff between performance and security. It is also possible to use the OT extension technique presented in \cite{OTexp} to increase the efficiency of oblivious transfer.

Although we use only image files as documents in our experimental implementation, we stress that the same mechanism can be used for all types of data for which robust watermarking schemes exist.

%% file: scenarios.tex
\begin{figure}[!t]
\centering
  \setlength\fboxsep{8pt}
  \setlength\fboxrule{0.5pt}
	\includegraphics[width=.5\textwidth]{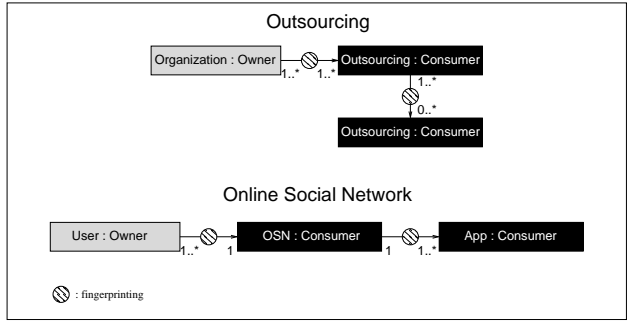}
	\caption{Outsourcing Scenario}
	\label{fig:all}
\end{figure}

\section{Scenarios}
\label{sec:scenarios}

\subsection{Outsourcing} 

The first diagram in Figure~\ref{fig:all} shows a typical outsourcing scenario. An organization acts as owner and can outsource tasks to outsourcing companies which act as consumers in our model. It is possible that the outsourcing companies receive sensitive data to work on and as the outsourcing companies are not necessarily trusted by the organization, fingerprinting is used on transferred documents. The outsourcing company itself can outsource tasks to other outsourcing companies and thus relay the documents, again using fingerprinting.

If now at any point one of the involved outsourcing companies leaks a confidential document, the organization can invoke the auditor to find the responsible party. The auditor then examines the fingerprints in the document, creates a lineage and is finally able to name the guilty party.
In the example given in the introduction~\cite{offshoreOutsourcing}, there were three outsourcing companies involved and a data leakage could not be clearly associated with one of these. Using \sysname\ the responsible party can be clearly found.

\subsection{Online Social Network}

The second diagram in Figure~\ref{fig:all} shows an online social networking scenario. The users of the network are the owners, as they enter their personal information, post messages, etc. The online social network (OSN) uses all this information as a consumer in this scenario. Third party applications that have access to this information in return for some service act as further consumers in this scenario. The users give their information to the OSN which can relay that information to third party applications using fingerprinting. 

In the introduction we gave an example where third party applications leaked private information of Facebook users to advertising companies~\cite{FacebookAppLeak}. Although it was possible to determine which applications leaked data, given some leaked information it is not possible to find the responsible application. Using \sysname\ this link can be found and proven.

%% file: discussion.tex
\section{Discussion}
\label{discussion}
\subsection{Composability}
\label{composability}
\sysname\ also allows us to create a lineage for a document that is published as part of a composed object. Consider the following setting: Owner~A and Owner~B each own a set of database entries $D_A$ and $D_B$ which they both transfer to Consumer~1, who receives marked versions $D_A^1$ and $D_B^1$. Consumer~1 then creates a composed object $D^1 = D_A^1 || D_B^1$ by concatenation and transfers it further to Consumer~2, who receives a marked version $D^2=D_A^2||D_B^2$ and then leaks it. If Owner~1 notices that his set of database entries was published as part of the composed object $D^2 = D_A^2||D_B^2$, he invokes the auditor and provides him with $D^2$, $D_A^2$, $D_A$, the watermark and the necessary detection keys. The auditor detects the watermarks in $D_A^2$ and verifies that this version was transferred to Consumer~1, who proves that he transferred the composed data to Consumer~2 by showing the watermark and the detection keys for the composed document. As Consumer~2 cannot disprove his guilt, he can be held accountable for publishing a version of $D_A$ even though he published it as part of composed data.

\subsection{Collusion Resistance}
The collusion resistance of our scheme depends on the collusion resistance of the underlying watermarking scheme. Assume several consumers are working together in order to create an untraceable version of a document. Then their approach is to merge the versions they rightfully obtained to create a new version where the watermarks cannot be detected. As the detection of $\sigma$ is just a detection of a watermark in the complete document, we obviously have the same collusion resistance as the watermarking scheme for this case. 
The case of the detection of a bit $b_i$ in a part $D_i$ is again just a detection of a watermark, so the collusion resistance is again the same as for the watermarking scheme. However, we have to know which detected bit belongs to which consumer, so that we can still guarantee that the sender cannot frame the receiving consumers. Linking the detected bits to the responsible consumers is possible, as for each consumer a different embedding key was used.
As for each part multiple bits might be detectable, the probability for a sender to successfully frame the receiving consumers is less than or equal to the probability of framing a single recipient successfully, as he still would have to guess all the bits correctly.
However, we have to note that in order to successfully mount a collusion attack against our scheme, it is sufficient to mount a collusion attack against 1 of the $n + 1$ watermarks that are used, where $n$ is the number of parts the document was split into. 

We can conclude that our scheme tolerates collusions to a certain extent, when it is used with a collusion resistant watermark, without losing its key properties.

\subsection{Error Tolerance}
Depending on the quality of the underlying watermarking scheme, it may be too strong to require that all bits $b_i$ are detected correctly. Therefore, it could be a good idea to introduce some error tolerance. However, we have to keep in mind that this will increase the probability of the sender successfully framing an innocent recipient. There are two different kinds of errors that can occur: the first one is that no bit can be detected, and the second one is that a wrong bit is detected. Assume the document is split into $n$ parts. Tolerating a non-detectable bit increases the probability of successful framing by a factor of 2. Instead of guessing a bitstring $\overline{b} \in \{0,1\}^n$, it is sufficient to guess $\overline{b} \in \{0,1\}^{n-1}$. 
Tolerating a wrong bit is worse, as it increases this probability by a factor of $(n+1)$. Instead of accepting just the correct bitstring, we also accept all bitstrings that are changed at exactly one position. As there are $n$ positions, we additionally accept $n$ bitstrings; hence the number of accepted bitstrings and thus the probability of guessing one of these is higher by a factor of $n+1$. 
If we want to allow some error tolerance while keeping the probability of successful framing to be small, we have to choose a larger $n$; e.g., to tolerate 128 non-detectable bits, we choose $n=256$ and have the same framing probability as with $n=128$ and no tolerance.

\subsection{Possible Data Distortion}
In our experiment, we used a simple splitting algorithm: We split the image into $n$ equally sized squares. However, when we used a strong watermark for the small parts (that is the $\alpha$-factor used by the Cox algorithm is 0.5), differences between adjacent parts became visible even though the single watermarks are imperceptible. 
The resulting image can be seen in Figure~\ref{lena_scrambled} in the appendix. 
In some cases, this distortion might affect the usability of the document. We stress however, that we were still able to obtain good results with our approach. 
In Figure~\ref{lena_non_scrambled} we used the Cox algorithm with an alpha factor of 0.1 and no distortion is visible.

It might be interesting to investigate if this problem can be circumvented by using more elaborate splitting algorithms. As most watermarking schemes make use of the contiguity of information in the document, this is not a trivial task.

%% file: relwork.tex
\section{Related Work}\label{sec:relwork}

\subsection{Other Models}
The model introduced in \cite{papadimitriou2011data} intends to help the data distributor to identify the malicious agent which leaked the information. In addition, they argue that current watermarking techniques are not practical, as they may embed extra information which could affect agents' work and their level of robustness may be inadequate. In \sysname\ the relationship of data distributor and agents corresponds to the relationship between data owner and consumer and the model could be used as an alternative method to trace the information given to the consumers.

Controlled data disclosure is a well-studied problem in the security literature,
where it is addressed using access control mechanisms.
Although these mechanisms can control release of confidential information 
and also prevent accidental or malicious destruction of information,
they do not cover propagation of information by a recipient 
that is supposed to keep the information private.
For example, once an individual allows a third party app to access her information from a social network,
she can no longer control how that app may redistribute  the information.
Therefore, the prevalent access control mechanisms are not adequate 
to resolve the problem of information leakages.
Data usage control enforcement systems~\cite{DataUsageControl08,DataUsageControl13} employ preventive measures
to ensure that data is transferred in distributed systems in a controlled manner preserving 
the well-defined policies. Techniques have been developed for securely distributing data by forming coalitions among 
the data owners~\cite{OwnerCoaltion}.
In controlled environments, 
such techniques can be composed with our protocols to improve data privacy.

In \cite{Insider} the authors present the problem of an insider attack, where the data generator consists of multiple single entities and one of these publishes a version of the document. Usually methods for proof-of-ownership or fingerprinting are only applied after completion of the generating process, so all entities involved in the generation process have access to the original document and could possibly publish it without giving credit to the other authors, or also leak the document without being tracked. As presented in the paper, this problem can be solved by the usage of watermarking and possibly even by using complete fingerprinting protocols during the generating phase of the document.

\subsection{Other Fingerprinting Protocols}
In \cite{poh2009design} Poh addresses the problem of accountable data transfer with untrusted senders using the term \emph{fair content tracing}. He presents a general framework to compare different approaches and splits protocols into four categories depending on their utilization of trusted third parties, i.e., no trusted third parties, offline trusted third parties, online trusted third parties and trusted hardware. Furthermore, he introduces the additional properties of recipient anonymity and fairness in association with payment. All presented schemes use watermarking to trace the guilty party and most presented protocols make use of \emph{watermarking in the encrypted domain}, where encrypted watermarks are embedded in encrypted documents~\cite{WmEncDom}.
A new scheme presented is based on chameleon encryption~\cite{anderson1997chameleon}.
In \cite{FingerprintingSurvey} Sadeghi also examines several fingerprinting schemes and presents new constructions for symmetric, asymmetric and anonymous fingerprinting schemes. The asymmetric scheme uses a homomorphic commitment scheme to compute the fingerprinted version of the document.

Domingo-Ferrer presents the first fingerprinting protocol that makes use of oblivious transfer in \cite{domingo1999anonymous}. In the scheme, documents are split into smaller parts and for each part two different versions are created. Then the recipient receives one version of each part via oblivious transfer and in return sends a commitment on the received part. The recipient can now be identified by the unique combinations of versions he received. The protocol has several flaws, as discussed in \cite{sadeghi2001break} and \cite{hanaoka2005break}.  The main problem is that a malicious sender can offer the same version twice in the oblivious transfer, so that he will know which version the recipient receives. 

Sadeghi~\cite{sadeghi2001break} and Hanaoka et al.~\cite{hanaoka2005break} propose different solutions; the former lets the sender open some pairs to validate that they are not equal and the latter uses oblivious transfer with a two-lock cryptosystem where the recipient can compare both versions in encrypted form. However, both proposed solutions have some flaws themselves. The problem is that it is possible to create two different versions with the same watermark, so even if the equality test fails, the two offered versions can still have the same watermark and the sender will know which watermark the recipient received. Also, the fix proposed in \cite{hanaoka2005break} ruins the negligible probability of failure, as it does not split the document into parts, but creates $n$ different versions and sends them via 1-out-of-n oblivious transfer. 

Domingo-Ferrer presents another protocol based on oblivious transfer in \cite{domingo2000efficient}, but again the sender can cheat during oblivious transfer.
\cite{hu2010asymmetric} presents another protocol using oblivious transfer. The protocol uses an approach similar to the chameleon encryption~\cite{anderson1997chameleon}, and using 1-out-of-n oblivious transfer a decryption key is transmitted so that the sender does not know it. The protocol suffers from the same problems as the one presented in \cite{hanaoka2005break}; namely, the sender can guess the key used by the recipient with non-negligible probability $\frac{1}{n}$ and the sender can even cheat in the oblivious transfer by offering the same key $n$ times, so that he will know the key used by the recipient.

We see that all asymmetric fingerprinting protocols based on oblivious transfer that have been proposed so far suffer from the same weakness. We circumvent this problem in our protocol by additionally sending a signed message including the watermark's content, so that the recipient is able to prove what he asked for. In contrast to the watermark, this message can be read by the recipient, so he can notice if the sender cheats.

\subsection{Broadcasting}
\label{broadcasting}
Parviainen and Parnes present an approach for distributing data in a multicast system, so that every recipient holds a differently watermarked version~\cite{parviainen2001large}. The sender splits the file into blocks and for each block he creates two different versions by watermarking them with different watermarks and encrypting them with different keys. Each recipient is assigned a set of keys, so that he can decrypt exactly one version of each part. The resulting combination of parts can uniquely identify the recipient.    
In \cite{Fingercasting} Adelsbach, Huber and Sadeghi  show another approach for a broadcasting system that allows identification of recipients by their received files. With a technique called \emph{fingercasting}, recipients automatically embed a watermark in files during the decryption process. The process is based on the chameleon cipher~\cite{anderson1997chameleon}, which allows one to decrypt an encrypted file with different decryption keys, to introduce some noise that can be used as a means of identification. 
In \cite{KSUS07} Katzenbeisser et al. use the technique of fingercasting together with a randomized fingerprinting code in order to provide better security against colluding attackers.
However, in these broadcasting approaches the problem of an untrusted sender is not addressed.

\subsection{Watermarking}
\label{app_wm}
\sysname\ can be used with any type of data for which watermarking schemes exist. Therefore, we briefly describe different watermarking techniques for different data types.
Most watermarking schemes are designed for multimedia files such as images~\cite{potdar2005survey}, videos~\cite{doerr2003guide}, and audio files~\cite{alsalami2003digital}. In these multimedia files, watermarks are usually embedded by using a transformed representation (e.g. discrete cosine, wavelet or Fourier transform) and modifying transform domain coefficients.

Watermarking techniques have also been developed for other data types such as relational databases, text files and even Android apps. The first two are especially interesting, as they allow us to apply \sysname\ to user databases or medical records.
Watermarking relational databases can be done in different ways. The most common solutions are to embed information in noise-tolerant attributes of the entries or to create fake database entries~\cite{halder2010watermarking}.
For watermarking of texts, there are two main approaches. The first one embeds information by changing the text's appearance (e.g. changing distance between words and lines) in a way that is imperceptible to humans~\cite{brassil1999copyright}. The second approach is also referred to as language watermarking and works on the semantic level of the text rather than on its appearance~\cite{ARHKSTT02}.
A Mechanism also has been proposed to insert watermarks to Android apps~\cite{zhou2013appink}.
This mechanism encodes a watermark in a permutation graph and hides the graph as a linked list in the application. Due to the list representation, watermarks are encoded in the execution state of the application rather than in its syntax, which makes it robust against attacks.

Embedding multiple watermarks into a single document has been discussed in literature and there are different techniques available \cite{MultiWatermarking,MSU06,frikken2003cropping}. In \cite{MSU06} they discuss multiple re-watermarking and in \cite{frikken2003cropping} the focus is on segmented watermarking. Both papers show in experimental results that multiple watermarking is possible which is very important for our scheme, as it allows us to create a lineage over multiple levels.

It would be desirable not to reveal the private watermarking key to the auditor during the auditor's investigation, so that it can be safely reused, but as discussed in \cite{cox1997public,linnartz1998analysis,wu2012pollution} current public key watermarking schemes are not secure and it is doubtful if it is possible to design one that is secure.
In \cite{Sad08} Sadeghi presents approaches to zero-knowledge watermark detection. With this technology it is possible to convince another party of the presence of a watermark in a document without giving any information about the detection key or the watermark itself. However, the scheme discussed in \cite{Sad08} also hides the content of the watermark itself and are therefore unfit for our case, as the auditor has to know the watermark to identify the guilty person. Furthermore, using a technology like this would come with additional constraints for the chosen watermarking scheme.

%% file: conclusion.tex
\section{Conclusion and Future Directions}\label{sec:conclusion}
We present \sysname, a model for accountable data transfer across multiple entities. We define participating parties, their inter-relationships and give a concrete instantiation for a data transfer protocol using a novel combination of oblivious transfer, robust watermarking and digital signatures. We prove its correctness and show that it is realizable by giving microbenchmarking results. By presenting a general applicable framework, we introduce accountability as early as in the design phase of a data transfer infrastructure.

Although \sysname\ does not actively prevent data leakage, it introduces reactive accountability. Thus, it will deter malicious parties from leaking private documents and will encourage honest (but careless) parties to provide the required protection for sensitive data.
\sysname\ is flexible as we differentiate between trusted senders (usually owners) and untrusted senders (usually consumers). In the case of the trusted sender, a very simple protocol with little overhead is possible. The untrusted sender requires a more complicated protocol, but the results are not based on trust assumptions and therefore they should be able to convince a neutral entity (e.g. a judge). 

Our work also motivates further research on data leakage detection techniques for various document types and scenarios.
For example, it will be an interesting future research direction to design a verifiable lineage protocol for derived data.